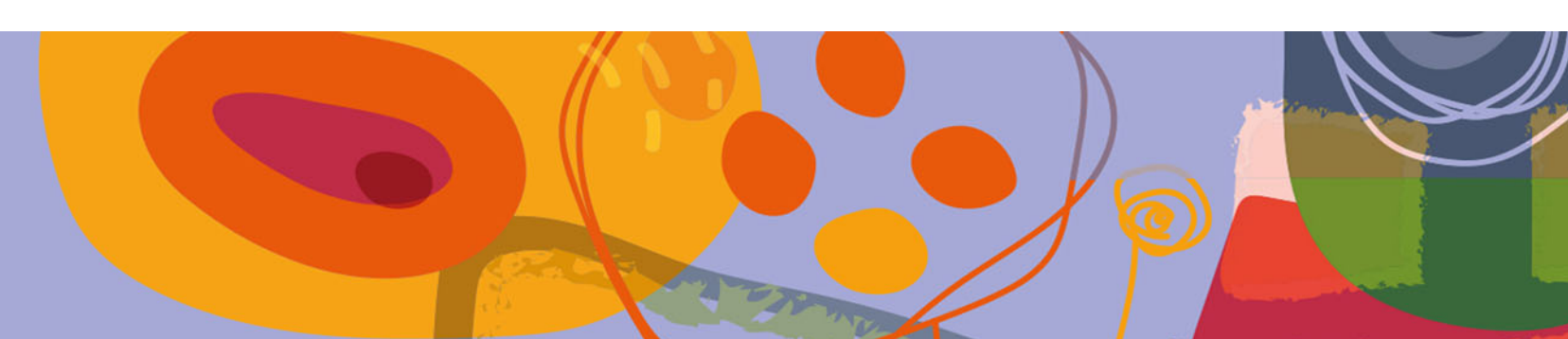

ARTICLE

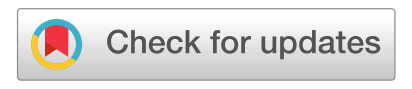

# Women in science: measuring participation in Europe across disciplines, generations and over time

Marek Kwiek[1]✉ & Lukasz Szymula[2]

In this research, we quantify an inflow of women into science in the past three decades. Structured Big Data allow us to estimate the contribution of women scientists to the growth of science by disciplines (N = 14 STEMM disciplines) and over time (1990–2023). A monolithic segment of STEMM science emerges from this research as divided between the disciplines in which women's share rose most rapidly–and the disciplines in which the role of women was marginal. There are four disciplines in which 50% of currently publishing scientists are women; and five disciplines in which more than 50% of currently young scientists are women. But there is also a cluster of four highly mathematized disciplines (MATH, COMP, PHYS, and ENG) in which the growth of science is only marginally driven by women. Digital traces left by scientists in their publications indexed in global datasets open two new dimensions in large-scale academic profession studies: time and gender. The growth of science in Europe was accompanied by growth in the number of women scientists, but with powerful cross-disciplinary and cross-generational differentiations. We examined the share of women scientists coming from ten different age cohorts for 32 European and four comparator countries (the USA, Canada, Australia, and Japan). Our study sample was $N = 1{,}740{,}985$ scientists (including 39.40% women scientists). Three critical methodological challenges of using structured Big Data of the bibliometric type were discussed: gender determination, academic age determination, and discipline determination.

[1] Center for Public Policy Studies, Adam Mickiewicz University in Poznań, Poznań, Poland. [2] Department of Artificial Intelligence, Adam Mickiewicz University in Poznań, Poznań, Poland. ✉email: kwiekm@amu.edu.pl

## Introduction

Derek de Solla Price (1963) presented his hypothesis of growth of scientific knowledge according to which the growth would shift from exponential to a steady state of production. However, scientific publishing was never saturated, as expected (Wagner and Kim, 2014). Big Science was replaced with global mega-science (Baker and Powell, 2024), with world science experiencing exponential growth (Larivière et al., 2008). One missing link in Price's hypothesis was women in science. In the early 1960s, the share of women in academic science was still marginal. However, recent growth of global science in general (Bornmann et al., 2021) is accompanied by the powerful growth of women in science. In the present research, we estimate the scale of women's inflow to academic science in Europe over the past three decades by using structured Big Data of the bibliometric type.

Although new big datasets clearly open new possibilities (Liu et al., 2023; Sanliturk et al., 2023; Sugimoto and Larivière, 2023), their limitations in examining the changing role of women in science from a historical perspective require consistent rethinking along methodological lines. We offer a discussion of these limitations using the growth of science in Europe in 1990–2023 as seen through the lenses of the growth of men and women scientists.

Our focus is on the methodological choices to be made while using structured, curated, commercial Big Data of the bibliometric type (Holmes, 2017; Selwyn, 2019; Salganik, 2018), the resulting limitations of using bibliometric data in women-focused research designs, and the implications for the studies of growth of science from a gender perspective.

From an empirical perspective, we examine the changing share of women scientists of different ages (coming from different age cohorts) historically (in the 1990–2023 period) and within 14 science, technology, engineering, mathematics, and medicine (STEMM) disciplines for 32 European and four comparator countries (the USA, Canada, Australia, and Japan).

From a historical perspective, until the early 2010s, large-scale quantitative science studies still referred to scientists in general (with no gender used) rather than to men and women scientists. Bibliometric datasets (including the two largest and competing commercial datasets: Web of Science owned by Clarivate Analytics and Scopus owned by Elsevier) with publication and citation metadata on dozens of millions of publications until the early 2010s were not used to study women's contributions to the academic science enterprise. The major focus in large-scale quantitative science studies using bibliometric datasets was on publications because the available datasets were constructed to be used in studies of publications rather than of scientists. Only gradually since the early 2010s has the publication as the unit of analysis been complemented by the scientist (increasingly characterized by gender) as the unit of analysis.

In contrast, small-scale, survey-based, and interview-based research was used extensively to examine women in science (see Baker, 2012; Cole, 1979; Fox and Mohapatra, 2007; Long et al., 1993; Rosser, 2004; Sonnert, 1995; Zuckerman et al., 1991), with dozens of influential studies offering such important hypotheses as leaky "pipeline" and "pathways," "chilly climate," and different "climates" in academic science departments, more or less welcoming women in the STEMM fields, "accumulative disadvantage" in women's careers, "leaving science," and the "Matilda effect", "glass ceiling effect", and "glass borders" in careers, "productivity puzzle" and many others (e.g., Branch and Alegria, 2016; Cornelius et al., 1988; Fox, 2010; Fox and Kline, 2016; Goulden et al., 2011; Shaw and Stanton, 2012; Xie and Shauman, 2003; Zippel, 2017).

Only later on, with the emergence of the first large-scale, gender-detecting tools (Karimi et al., 2016; Liu et al., 2023; Wais, 2016), could the growth of science be shown from large-scale data-driven *a*nd gender perspectives (Gui et al., 2019). Increasingly, women in science have been studied with ever more precision and strengthened methodological clarity. Although this expanding field is in the early stages of development, its policy impact may be substantial.

The growth of global networked science (Wagner, 2008) can be analyzed through a variety of methodologies; however, quantitative science studies are probably best equipped to explore the extent of the globalization of science and growth of science in spatial and temporal, individual and collective, national and cross-national dimensions using global publication and citation data. Global mega-science is being driven as much by collaboration as by competition among scientists (Powell, 2018; Kwiek and Roszka, 2021a). The global changes in how science is conducted are fundamental, and the accounts of these transformations are numerous (Marginson, 2020; Baker and Powell, 2024; Wagner, 2008; Wang and Barabàsi, 2021). In the current paper, we conceptualize and quantify global science as the totality of publications indexed in the global datasets, here in the Scopus database; analogously, the European science examined in the empirical part of the current paper is the totality of publications indexed in Scopus and authored by scientists with European affiliations. We follow Baker and Powell's assumption (2024: 10) that scientific journal articles published in peer-reviewed journals "is the most valid, historically consistent, and readily obtainable indicator of the volume of scientific inquiry at any one time, point, and place".

In this research, we pose three major research questions regarding an inflow of women scientists to STEMM disciplines: first, did women enter different STEMM disciplines with similar intensity in the past three decades (1990–2023); second, what is the current participation of women in STEMM disciplines by academic generations (based on publishing experience); and third, do the four highly mathematized disciplines (MATH, COMP, PHYS, and ENG) form a separate cluster in which access of women to the publishing academic profession is limited compared with the other disciplines. Our objective is to study these three research questions descriptively using micro-level data on scientists from 14 STEMM disciplines across 32 European countries.

## Women in science: traditional studies

Although traditional studies using survey and interview methodologies—and leading to the emergence of major hypotheses about the role of women in science—were using limited numbers of observations (e.g., Baker, 2012; Sonnert, 1995), current large-scale studies use hundreds of thousands and millions of observations to explore similar academic career phenomena (see, e.g., Ross et al. (2022) on credit allocation in science; Huang et al. (2020) on gender differences in publishing career lengths and drop-out rates; Ni et al. (2021) on gendered nature of authorship; Ioannidis et al. (2023) on gender imbalances among top-cited scientists; King et al. (2017) on gender disparities in self-citations as well as Larivière et al. (2013) and West et al. (2013) on gendered nature of authorships in science).

The new scale comes at a cost, however. The number of variables available to explore the phenomena under study substantially decreases when bibliometric datasets are used, limiting the power of regression-based approaches. We learn more about the patterns related to the participation of women in science, but the mechanisms at play are hard to explore quantitatively. We need to remember about using new technologies "wisely" rather than "irresponsibly" (Weingart, 2004). The vast majority of large-scale studies use only one source of data: the bibliometric dataset;

studies in which bibliometric datasets are combined with survey (or interview) data are rare, and the connection between these heterogeneous data sources is relatively weak.

Hence, although traditional small-scale studies have produced influential hypotheses and theories, current large-scale research is more descriptive than theoretical in nature. The combination of inspiring older small-scale theoretical inputs and recent large-scale descriptions may substantially push the frontier of research on women in science forward. Detailed empirical analyses of how science and scientists work can go beyond national borders: Traditionally, most studies on—and data about—women in science originated from the United States, arguably the largest and most studied science system in the world.

## The promise of digital traces, cohort, and longitudinal study designs

Digital traces left by scientists in their publications indexed in global datasets over their entire academic careers open a fundamentally new dimension in academic profession studies: the dimension of time. Firmly embedding academic work in temporality allows us to study gender disparities at the minute detail by following the same scientists over time (in longitudinal study designs) within the same disciplines.

Large-scale quantitative science studies and traditional small-scale studies from the sociology of science explore similar issues: the social stratification of science, accumulative advantage and disadvantage in academic careers, and gender disparities in academic work. However, the former provides analyses globally, regionally, or in the various aggregations of countries, leading to new cross-national insights. The globalization of science can be explored with new tools and on an unprecedented scale: men and women scientists can be studied within their disciplines (e.g., mathematicians and computer scientists can be contrasted with biochemists and microbiologists globally, as in Kwiek and Szymula, 2024; Kwiek and Szymula, 2025) or within their age cohorts globally so that young men scientists in a discipline can be contrasted not only with young women scientists but also scientists who were young in the early 2000s and those who were young in the early 2010s.

In other words, new cross-disciplinary studies can follow cohort study designs (exploring the various aspects of academic work and careers by successive, nonoverlapping cohorts of scientists or by scientists entering academic publishing and academic careers in different years); and these studies can follow longitudinal study designs (tracking the same individuals, men, and women scientists over their careers for decades, see Kwiek and Roszka, 2024a, Kwiek and Szymula, 2025).

This type of individual, micro-level data for every (publishing) scientist globally from a gender, cohort, and disciplinary perspective allows for testing and expanding the traditional theories developed in the sociology and economics of science. At the same time, however, the focus of large-scale, cross-national studies is mostly descriptive, and the data used are quantitative and individual in nature. At the global level, there is still limited access to variables and indicators related to research organizations (from universities to hospitals to corporations) and their units (such as departments within universities)—and limited access to individual traits—other than those derived from bibliometric datasets. In bibliometrics, there are individuals and their publication and citation metadata. There is no access to individuals' perceptions of the way things are done around them in their departments (departmental climates, cultures, or "ways of doing business" in science at the lowest organizational level). To analyze these dimensions of doing science globally, global surveys of the academic profession are needed.

## The global growth of science from a gender perspective

The global growth of science from a gender perspective is a research theme that could be explored large-scale only recently. When we look at the influential research on the global growth of science, women are not discussed in them at all. This is understandable considering there is no access to data on women in science at the micro level of individuals and only aggregated data available from global and regional statistics (OECD, UNESCO, and the European Union scientific workforce datasets). To give an example of the practical invisibility of women scientists in studies on the growth of science, the word "women" (or its alternate, "female") does not appear in numerous papers on the theme at all. Price (1963), who referred to women just once, is not an exception. The data on women in science were just not collected; consequently, female scientists were statistically invisible.

Women do not appear in Ben-David (the author of the influential *The Scientist's Role in Society*) in his extended study of the growth of science in Germany and the United States (Ben-David, 1968), which was published in *Minerva*; in Riesman (1969) in his correspondence to *Minerva* about Ben-David's study; in Gilbert (1978) in his review of indicators of scientific growth, which focused on growth in manpower (in our reading today: men and women) and growth in knowledge (in our reading today: publications); in Michels and Schmoch (2012) in their research linking the growth of science and database coverage or the rising numbers of older and newer journals indexing publications globally; in Wagner and Kim (2014) in their study revisiting Price's hypothesis of growth of scientific knowledge in the context of the proliferation of venues for publication; in Ioannidis et al. (2014) in their study of the "continuously publishing core" in the global scientific workforce, showing that only about 150,000 scientists (we would add: men and women scientists) out of 15 million have an uninterrupted, continuous presence in the Scopus database, having a publication every year in a 16-year period.

In their research into growth rates of modern science since the mid-1600s based on publications and cited references indexed in WoS, Bornmann and Mutz (2015) never mentioned women scientists. Zhang et al. (2015), in their research about the transformation of Price's "big science" into global mega-science and on the shifting global center of gravity of science production in the twentieth century, did not refer to women in science; also Bornmann et al. (2021) in their research into growth rates of modern science based on the two established datasets (WoS and Scopus) and two new datasets (Dimensions from Digital Science and Microsoft Academic) did not mention women. The so-called Stanford List of the upper 2% of highly cited researchers globally, which is prepared annually by John Ioannidis et al. and Elsevier (Ioannidis et al., 2014) and which is based on solid methodology and objective indicators, has not indicated the gender of its 200,000 most highly influential scientists. Finally, Baker and Powell in their recent landmark study of the growth of "global mega-science" (2024) focus on scientists in general, and their publications.

## Women's invisibility in large-scale data and its implications

What does this list of influential research on growth in science—without references to women in science—indicate? There are two straightforward implications of women's theoretical and practical invisibility in science growth literature: First, in large-scale quantitative science studies and large-scale studies in sociology of science and bibliometrics in the past half a century, when women in science did not exist in the data, they were not present in the discussions. Speculations without empirical support would not be pertinent to follow. In contrast, small-scale survey-based, and interview-based research abounded in the very same half century. However, the theories and hypotheses produced in this small-scale research have not entered the much wider studies of science, even in parentheses. Second, now that access to micro-data in global datasets is ever wider, both to commercial datasets of the Scopus and Web of Science type (as well as Academic Analytics with the USA data) and new noncommercial datasets of the OpenAlex type (Priem et al., 2022), the role of women in the growth of science might be examined from global and historical perspectives.

## New datasets: data analysis and reflection needed

The raw data at the micro level of individual scientists are ever more available. The assignment of scientists in general to men and women scientists in particular is becoming ever more technically possible (in binary terms, a recent survey-based study of the nonbinary representation in 21 professional STEM societies reports nonbinary scientists to be at the level of 0.56%; Cech and Waidzunas, 2021). Long years of research within the tradition of scientometrics and bibliometrics have provided a good starting point in research, and reliable and easy-to-use gender detection tools (as discussed below) make large-scale men/women analyses possible. Finally, the scale of participation of women in science is phenomenal (in STEMM fields, i.e., far beyond the traditional broad fields of social sciences and humanities and the traditional disciplines of education, psychology, and linguistics)—which urges data analysis and reflection, both retrospectively and about the status quo in science today.

More data analysis and reflection are needed when we realize that women enter traditionally male-dominated disciplines in large numbers and high percentages. In some STEMM disciplines, women are becoming a majority (or coming close to majority) of currently publishing scientists (as in 2023, according to our computations for Europe in IMMU immunology and PHARM pharmacology, toxicology, and pharmaceutics, and coming closely to being a majority in MED medicine, BIO biochemistry, genetics, and molecular biology as well as NEURO neuroscience). The double theme of women's participation in STEMM and their attrition in STEMM becomes more important than ever before, and collective refinement of methodologies that can describe the scale of ongoing transformations of science from a gender perspective is becoming ever more urgent.

From an empirical perspective, as our computations based on raw Scopus data presented in the paper show, the growth of science in the past few decades by gender was not spread evenly across the STEMM disciplines. There have been powerful differences between the growth of women in highly mathematized STEMM disciplines (e.g., MATH, COMP, PHYS, and ENG in terms of the Scopus journal classification codes, All Science Journal Classification, ASJC) and the growth of women in the rest of STEMM areas (especially in MED and BIO). In 2023, young women in the European systems studied already outnumbered young men in six STEMM disciplines—but they were lagging behind in increasing their shares in the four highly mathematized disciplines.

## Measuring the growth of science and women: three major methodological challenges

In the current study, we examine the share of women scientists of different ages (coming from different age cohorts) historically (in the 1990–2023 period) and within 14 STEMM disciplines for 32 European and four comparator countries. As a result, major challenges in using raw structured Big Data of the Scopus type are related to the nonambiguous determination of gender, age, and academic disciplines.

## The first challenge: gender determination

Scientists' gender does not appear directly in publications, but it can be defined on the basis of several variables that appear in publications. The only data source to determine the gender of a scientist in research on a global scale—in contrast to single nations' scale in which national registries of scientists may be available—is their publications. However, gender inferences from publication data are subject to limitations.

In short, bibliometric databases are not perfect and were not historically designed for the purpose of researching the scientific profession or for any other research topic (hence repurposing of this type of data for research purposes is necessary; see Salganik, 2018). The standard focus of bibliometric databases is the publication; similarly, the traditional unit of analysis in large-scale research on science is the publication, not the author.

There is a fundamental difference between the three major ways in which scientists' gender can be assigned.

(1) *Administratively assigned gender* in national registries of scientists and other official datasets listing the national science workforce, with scientists and their national-level IDs.
(2) *Self-declared gender* assigned in survey research (traditionally binary, increasingly moving beyond a binary approach in the past decade) and in interview-based research.
(3) *Gender determined by gender detection tools* based on metadata from global publication datasets, with different degrees of accuracy for different countries.

In the present research, we will use the third approach to gender assignment.

There is also a stark contrast between perfect binary gender identification used in (some) countries and some national datasets (Italy, Norway, Poland, and the USA being prime examples where a straightforward, administrative, and binary distinction is made between men and women scientists), on the one hand, and the way gender is used in cross-national or global research as in our study. Thus, administratively assigned gender (probability 100%) is in stark contrast with statistically derived gender (probability at least 85% in the gender determination tools we use). Men and women scientists, as shown in national statistics (with no error), are in stark contrast with men and women scientists as indicated by large-scale gender detection tools (with some margin of error).

Exploring the contribution of women to the growth of science over the past three decades requires one key determination: the unambiguous assignment of scientists to one of the two groups, either men or women scientists. In addition, scientists need to be unambiguously assigned to a scientific discipline, country of affiliation, and age cohort (based on the year of the first indexed publication in the Scopus database).

The main challenge of large-scale research carried out using individual researchers (rather than individual publications) is precisely the gender of the researcher. Studies of scientific growth based on increasing numbers of (indexed) publications by country, discipline, and over time seem to be less methodologically and technically complex because these metrics form the core of the large-scale, curated, commercial datasets of the WoS or Scopus type. More complicated are studies based on the increasing number of unique researchers (as authors of publications) by country, discipline, and over time in these databases. Still, more methodologically complex are those studies carried out using the increasing numbers of unique scientists by country, discipline, over time, and by gender. Our research addresses the growth of science from the latter perspective by adding two additional dimensions to the data portfolio of each scientist: academic age and a single, dominant scientific discipline.

Determining a scientist's gender on a massive scale today is only possible using the digital traces left in publications indexed in global bibliometric databases. Although in single-country studies, gender may be available for research purposes (e.g., the administratively provided gender from national registries of scientists), in global studies, special gender detection tools need to be used, and these have been available for a little over a decade.

Limitations of gender inference based on bibliometric databases include the following points: Not all publications indexed in these databases have full author names available (which is key to gender identification by gender detection tools), especially for older publications and multiauthor publications in disciplines such as astronomy; the proportion of publications with full names varies based on the scientific discipline, the period studied, and the country; there are different names that are not gender specific in different countries (examples of such unisex names include Andrea, Claude or Dominique; Sebo 2021). In addition, the gender detection tools currently in service are characterized by varying degrees of accuracy, depending, among other things, on the country studied as well as varying costs of accessing this service. Most tools are typically commercial in nature, but there are also tools available for free or for a small fee.

Ethical issues need to be raised: The concepts of sex and gender are not interchangeable (with biological aspects vs. sociocultural roles at play), and gender detection tools use a binary classification of gender. Nonbinary identification is not possible using large-scale bibliometric datasets (Peters and Norton, 2018), even though it may be available in self-reported identifications in academic profession surveys and interviews, and through manually browsing the content of the institutional web pages of scientists.

### The challenge of gender determination and gender detection tools

Before the first gender detection tools appeared in the research on science and scientists, gender determination could be conducted only manually. Manual gender determination, however, was not a scalable procedure, and it was based on the knowledge of first names from a given country (Santamaria and Mihaljević, 2018). Manual gender determination could also be based on social security administration baby names or on such unsupervised approaches to gender identification as mixed methods combining name-based detection methods with an image-based face recognition approach (Karimi et al., 2016).

Karimi et al. (2016) compared several detection tools available using a dataset of just over 1400 scientists with their CVs, websites, and face pictures. Introducing the R package *genderizeR* based on the *genderize.io* gender detection tool, Wais (2016) provided a comparison of the methods used in two influential large-scale studies of women in science (Larivière et al., 2013; West et al., 2013). Supervised learning methods of gender detection were developed and compared in Hu et al. (2021), who trained machine learning models on a dataset of registered users of an internet company (Verizon Media) with 21 million unique gender/name pairs and evaluated their models on that dataset combined with a social security administration baby names dataset with 98,400 unique first name/gender pairs. Recently, large language models (LLMs) have been tested to verify their gender detection potential, with several studies indicating that ChatGPT can outperform other tools (Alexopoulos et al., 2023).

Two major error types in the estimation of the changing proportions of women in science are measurement error and sampling error (see Science-Metrics 2018: 22–28). First, the inaccuracy of the approach to determine the gender of each scientist in our European sample causes measurement error. In our research, we use NamSor, one of the most reliable gender detection tools, which has a built-in measure of accuracy in gender determination: The results from the NamSor API provide the most likely gender for a given combination of first and last names, and they provide statistics of this attribution. Second, the proportion of women in the population of scientists is estimated on the sample of scientists for which gender can be inferred. Thus, our sample of European scientists is not picked randomly (it is a nonprobability sample because the inclusion in the sample is determined by the availability of the full name and capacity of NamSor to identify the gender from this name). Therefore, there is a risk that the sample used is not representative of the whole population of European scientists.

However, for most countries, the percentage of scientists with names of ambiguous gender is low, so the impact of the potential sampling error is low. In the present paper, following Science-Metrics (2018), we assume that this bias is negligible and that our sample is representative of the authors with ambiguous names. A full validation (based on stratified random sampling) with manual curation of several thousand individual scientists from different countries, disciplines, and age cohorts would be useful; however, it exceeds the possibilities of our research team.

Our study followed the approach used in Elsevier's recent reports on the role of women in science (Elsevier, 2018; Elsevier, 2020). Data on the specified gender of scientists were made available to us by the International Center for the Study of Research (ICSR) Lab. We used a dataset that contained the researcher's identifier in the Scopus database and their gender, as specified with a probability score of 85%. The dataset we used contained the author's identifier from the Scopus database and two variables determined using the NamSor gender inference tool. NamSor offers a high degree of accuracy (i.e., there are few false positives) and recall (i.e., there are few unknowns) as well as global coverage. Its validation procedure relies on the use of directories' listing names and geographical locations (Science-Metrix, 2018; NamSor, 2024). The NamSor software used for gender detection in our dataset has been positively evaluated in numerous studies (e.g., Santamaria and Mihaljević, 2018; Sebo, 2021).

### The second challenge: academic age determination

There is also a fundamental difference between the three types of age:

(1) *Administratively assigned year of birth* in national registries of scientists, here in accordance with personal data, straightforwardly leads to biological age determination at any point in time.
(2) *Self-declared biological age* in survey-based and interview-based research, with some margin of error.
(3) (Academic) *age determined using global publication datasets* available only through a proxy of the date of the first publication indexed in the dataset, with different degrees of accuracy for different countries and disciplines.

In the present research, we will use the third approach to age assignment.

Age is important for numerous dimensions of academic careers, including research productivity patterns, publishing patterns, and migration and international collaboration patterns as traditional studies of the academic profession show (Stephan and Levin, 1992) and recent studies from the economics of science keep reminding (Stephan, 2012).

The idea of a rigorously determined academic age works very well in the context of mature science systems, whose scientists have been present in global scientific journals for decades; and it does not work as well for newcomers to the global science enterprise, with increasing presence in the datasets in the past one or two decades only. Thus, there is a stark contrast between perfect age identification used in (some) countries and some national registries of scientists, on the one hand, and the way (academic) age as a proxy of biological age is used in our research and many bibliometric-based studies, on the other hand. An administratively assigned year of birth (with no error) is in contrast to (academic) age derived from the metadata of the first indexed publication. Academic age determination, which has different levels of accuracy for different systems, depends on the embeddedness of national systems in the global system of indexed publications.

The methodological issue of age determination is of critical importance for two types of analyses of (publishing) scientists, and Big Data make possible the following:

(1) *Cohort analysis* (Bell, 2020; Fosse and Winship, 2019; Glenn, 2005; O'Brien, 2015): Scientists starting publishing in different periods—or belonging to different publishing cohorts—can be compared.
(2) *Longitudinal analysis* (Menard, 2002; Ployhart and Vandenberg, 2010; Singer and Willett, 2003): Publishing scientists can be tracked over time, from their first publication indexed in a dataset onwards.

For instance, publishing, collaboration patterns, and mobility patterns for mathematicians (MATH, with a relatively small and slowly increasing share of women scientists over time) can be compared with the patterns for biochemists and molecular biologists (BIO, with a relatively large and still powerfully increasing share of women scientists over time), either globally or within selected aggregates of countries. Both cross-cohort and longitudinal analyses can be conducted at various granulation levels. Our empirical focus in the present research is on Europe, but the methodological issues explored in this section refer to other aggregation levels.

Academic age as a proxy of biological age in research into academic careers has been increasingly used since the early 2010s, when large-scale studies at the micro level of individuals started (e.g., Aksnes et al., 2011; Abramo et al., 2016; Milojević, 2018; Nane et al., 2017; Robinson-Garcia et al., 2020). For large-scale studies (as opposed to small-scale studies), data on biological age were not available for research purposes.

Biological age was used in studies of science and scientists for more than half a century, starting with a notable and widely criticized Lehman's book on *Age and Achievement* (1953); there is also a long line of research focused on the impact of productivity on social stratification in science, with data on age used, as in Pelz and Andrews (1976); Cole (1979); Kyvik (1990); and Stephan and Levin (1992). Biological age has also been widely used in research collaboration studies (Abramo et al., 2016; Rørstad et al., 2021; Kwiek and Roszka, 2021b).

Traditionally, two proxies have been used for biological age: first, academic age related to the date of the first publication (as in Nane et al., 2017; Radicchi and Castellano, 2013; Robinson-Garcia et al., 2020) and, second, academic age related to the date of receiving a PhD (as in Sugimoto et al., 2016; van den Besselaar and Sandström, 2016; Savage and Olejniczak, 2021). Although in national studies both options are viable, in cross-national studies, only the second option seems possible in practice. Changing the focus of research from a national to a multicountry or global, biological age becomes available only through a proxy for academic age.

A systematic analysis of the differences between academic age and actual biological age of a whole national system (Poland) shows that the level of correlation between the two types of age considerably differs between STEMM disciplines and social sciences and humanities (Kwiek and Roszka, 2022b). However, for the STEMM disciplines examined in the present research, the correlation coefficients are in the range of 80–90% (e.g., 0.89 for CHEM Chemistry. 0.88 for PHYS Physics and Astronomy, 0.85 for MATH Mathematics, and 0.90 for IMMU Immunology and Microbiology). The correlation issues may be bigger for scientifically "lagging" countries (not examined in this research) and smaller for "mature" and "developing" countries, to use Wagner's (2008: 88) typology of science systems.

The use of Scopus-based academic age as a proxy for biological age raises questions related to dataset bias. However, only bibliometric sources (and Scopus works here better than WoS) seem useful in large-scale studies exceeding national systems and using micro-level data from decades of academic publishing. The competing data sources would be the Google Scholar database, which has a bias in favor of younger scientists (see Radicchi and Castellano, 2013), and OpenAlex, which is still very difficult to be useful at the lowest level of individuals with unambiguously defined publishing and citation portfolios (see Alperin et al., 2024; Priem et al., 2022).

All data sources refer only to publishing scientists; by definition, all non-publishers are excluded from the analysis. Among many limitations of our research, perhaps the most salient is that the participation of women (and men) in the growth of science is viewed exclusively through the lenses of publishing scientists, with all other roles (teaching, administration, mentoring, etc.) being excluded.

### The third challenge: academic discipline determination

Finally, there are several ways to determine a single discipline for every individual scientist in a study sample. In national datasets, disciplines are usually ascribed to scientists based on the different national-level classifications that are often used in national research assessment exercises. Scientists tend to be ascribed to a discipline (lower order) and field (higher order).

A major challenge in moving from national studies to multicountry studies as ours is to find a common denominator in discipline lists for all the countries involved. Because seeking and coordinating 32 classifications of disciplines from 32 countries proved practically impossible, our methodological choice was to rely on the broader fields and disciplines used in the Scopus database. Therefore, all publications of all scientists in our sample were initially ascribed to journals, with their ASJC disciplinary codes. In addition, all publications in our sample had their lists of cited references. By combining all publications and cited references for every single scientist in our sample, we were able to determine a single dominant discipline for scientists.

To obtain the dominant discipline, a set of publications from the Scopus database was used. Publications were from 2023 and before and were restricted by source and type of publication: (a) journal article and (b) conference paper in a book or journal. Each cited reference from each publication was accompanied by its discipline as assigned by the discipline of the journal in which it appeared. The disciplines assigned to a cited reference were based on the four-digit ASJC code used by the Scopus database. To switch to a two-digit classification, unique disciplines were selected based on the first two digits of the four-digit value. Then, for each author, the number of cited references was counted for all disciplines referenced by the author (excluding the "multidisciplinary" discipline). For each author, the discipline with the highest number of cited references (the modal value) was selected. Authors who had more than one dominant discipline or no discipline were removed from the analysis.

The ascription of scientists to dominant disciplines tends to work better for scientists with longer publication lists and using more Scopus-cited references; for scientists with very short publishing careers and meeting only the minimum output requirement (five articles), the ascription might be more arbitrary.

The ascription to disciplines based on the minimum of 10 articles may be more accurate than when based on five articles; however, a large number of early-career researchers or scientists who stopped publishing early and quit science (Kwiek and Szymula, 2024) would be excluded from our study sample. In the present research, we do not apply ascribing scientists to different dominating disciplines over their different academic life periods, not to complicate the analysis (as we did in Kwiek and Roszka, 2024c with respect to the Polish system). We know that scientists do change their disciplines; however, in the current study, for the sake of simplicity, we use a “one scientist, one discipline” formula.

## Data and methods

**Data**. The major characteristics of the study sample for 1990–2023 (1,740,985 scientists, including 685,968, or 39.40%, women scientists) and for 2023 (684,155 scientists, including 275,204, or 40.23% women scientists) are shown in Supplementary Table 1 and Supplementary Table 2. Our sample (with all scientists meeting the initial criteria requirements) was constructed as follows: First, to determine the number of scientists, unique authors of publications who published their works from 1990 to 2023 were selected. For this selected group of authors, their publishing years were determined. The resulting set of scientists was then narrowed down according to a package of five restrictions: (a) affiliation in a European Union (EU-27) or in one of the five EU associated countries (the United Kingdom, Norway, Iceland, Switzerland, and Israel); (b) a STEMM dominant discipline; (c) gender (binary approach: man or woman); (d) a nonoccasional status in science: a minimum scientific output defined as five publications (Type: journal article, conference paper in a book or a journal) throughout the scientist’s career (lifetime); and (e) academic age, or the time passed since the first publication, here in the 1–50 years range. The four comparator countries (the USA, Canada, Australia, and Japan) have been selected based on their critical role in global science in the past three decades, as testified to by national publication output.

The minimum output in individual lifetime publication history allowed us to limit our sample to nonoccasional scientists. Generally, in terms of author name disambiguation, Scopus is reported as being more accurate than Web of Science (Sugimoto and Larivière, 2018: 36). Then, for each scientist, academic experience in full years, beginning in the year of the first publication of any type, was determined. For each year of a scientist’s research activities, the length of their academic experience and membership in the corresponding academic age group were determined. A sample for 1990–2023 was used for the trend analysis and a subsample for 2023 for a cross-sectional analysis. Figure 1 summarizes the sample’s design.

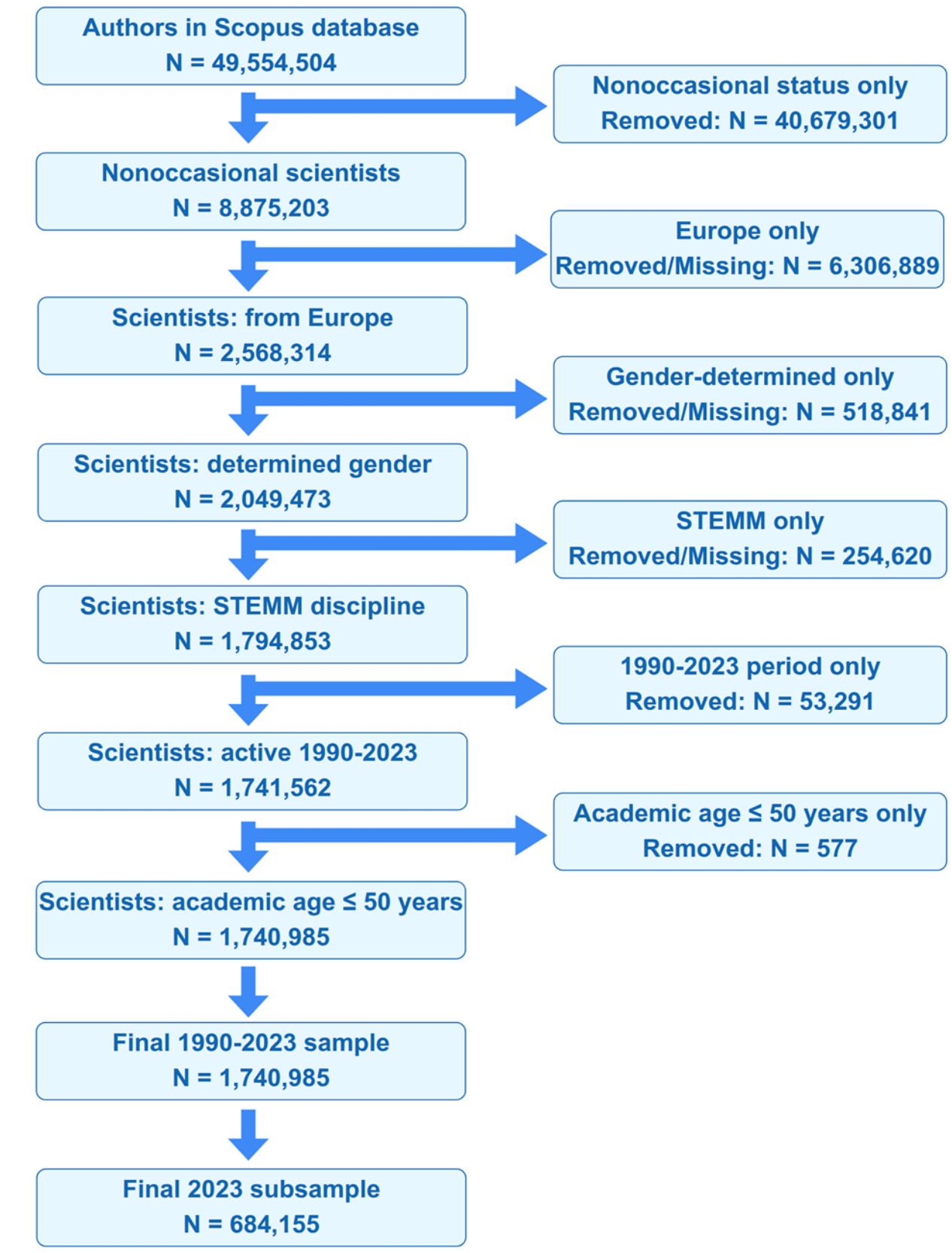

**Fig. 1** Flowchart: Stages in constructing the study sample and the 2023 subsample.

**Methods**. The raw data were made available to us by Elsevier under a multi-year agreement with the ICSR Lab. The Scopus database version for 2023 and before, dated March 29, 2024, was used. To obtain the results at the aggregate level, the operation in the ICSR Lab relied on the use of the Databricks environment, which allowed for managing and executing cloud computing with Amazon EC2 services. The scripts to generate the results were written using the PySparkSQL library. The operation was carried out on a 100% Scopus database with a snapshot date March 29, 2024, here using a cluster in standard mode with Databricks Runtime version 13.3 LTS with Apache Spark technology version 3.4.1, Scala 2.12, and an instance i3.2xlarge with 61 GB memory, eight cores, and one to four workers. The execution time for the entire script was 1.18 hours.

Our methodological approach is to estimate the growth of science from the combined perspective of gender and academic generations. Changing the proportions of women participating in science through publishing in the period of 1990–2023 (and separately in 2023) can be examined for various academic age groups. Academic age groups represent scientists with different types of research experience (as measured through a proxy of publishing experience: the years since the date of their first publication to 2023).

For each scientist in our sample, we have a set of unambiguously determined characteristics: gender, dominant country of affiliation, dominant academic discipline, total publication output (allowing us to define a nonoccasional status in academic knowledge production), and academic age (or academic experience, allowing us to define academic age cohort). Figure 1 shows the details of the sampling strategy: for instance, we had 2,568,314 scientists from Europe but we were able to determine gender of 2,049,473 scientists. Therefore, we removed from further analysis 518,841 observations (20.2%) for which gender was missing. We discuss briefly each characteristic in ESM.

**List of STEMM disciplines**. We focused on 14 STEMM disciplines, as defined by the journal classification system used in the Scopus database (All Science Journal Classification, ASJC): AGRI, agricultural and biological sciences; BIO, biochemistry, genetics, and molecular biology; CHEM, chemistry; COMP, computer science; EARTH, earth and planetary sciences; ENG, engineering; ENVIR, environmental science; IMMU, immunology and microbiology; MATER, materials science; MATH, mathematics; MED medicine, NEURO, neuroscience; PHARM, pharmacology, toxicology, and pharmaceutics; and PHYS, physics and astronomy.

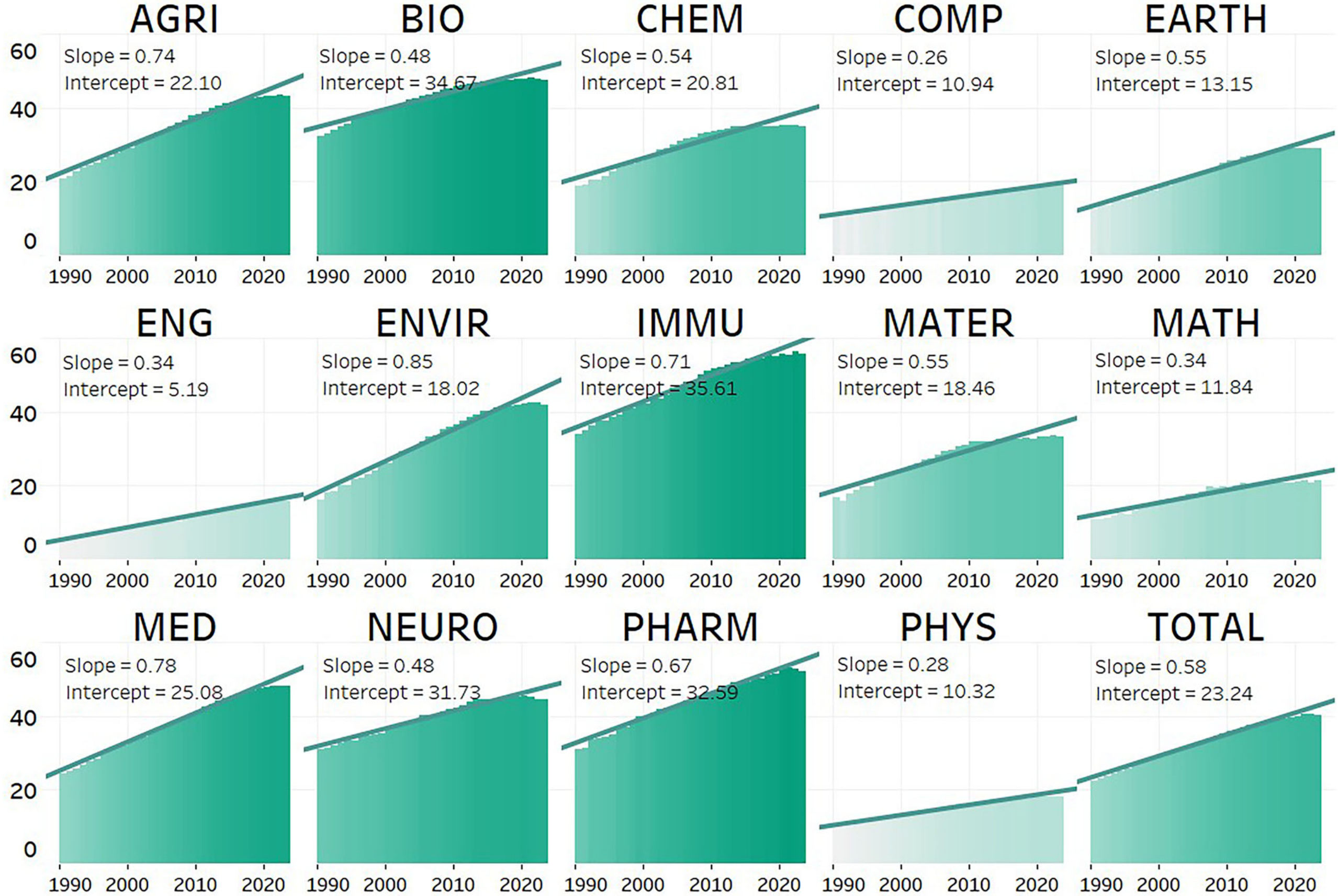

**Fig. 2** Growth in participation in publishing nonoccasional STEMM women scientists in European countries over time by discipline, 1990–2023 (in %) ($N = 1{,}740{,}985$).

## Results

**A trend view 1990–2023: the participation of women in science over time**. The growth of science in Europe in the past three decades was accompanied by phenomenal growth in the number of women scientists in the European science enterprise. However, powerful cross-disciplinary and cross-generational differentiations within the STEMM fields can be observed: The growth is not evenly distributed, with some disciplines being widely open to women and others still relatively closed, with slow or marginal changes in women's participation.

The trend of increasing participation of women in science is shown in Fig. 2, separately for 14 STEMM disciplines and for all disciplines combined (total). We used a linear trend ($y = at + b$, where $b$ is intercept, or the value where the trend line intersects the "y-axis" and $a$ denotes the slope of the trend line). The slopes, which are shown separately for each discipline, describe the steep trend lines. A slope of $a$ indicates the average change from year to year, and $b$ is the intercept indicating the level of the phenomenon—here: the percentage of women scientists among all research-active scientists—in the zero period (preceding the first year of analysis, i.e., in 1989).

Our data show that women's participation in science have different trends for different clusters of disciplines. In some disciplines, women's participation was already high in 1990, our starting point (reaching about one-third of all scientists publishing this year, as in BIO, IMMU, NEURO, and PHARM; see the details in Table 1), and reaching almost 50% in 2023. However, the fastest growth in the period was observed for MED (the largest discipline in Europe in terms of the number of scientists involved in publishing), AGRI, and ENVIR. In contrast, the cluster of four math-intensive disciplines had a low participation rate of women in the whole period examined and weak growth in 1990–2023: COMP, ENG, MATH, and PHYS. For all disciplines combined (total), women's participation almost doubled in the period (increasing from 22.44% in 1990 to 40.23% in 2023).

The percentage of female scientists has been steadily increasing in all disciplines studied, though at varying rates. The clusters of MATH, COMP, PHYS, and ENG had the lowest increase, as indicated by the slopes in regression model analysis being in the range of 0.26–0.34 (Table 2).

Each discipline showed a different time for a one-percentage point increase in the period studied. The fastest growth occurred for ENVIR (1.18 years), MED (1.28 years), and AGRI (1.35 years). The four disciplines of MATH, COMP, PHYS, and ENG were at the other extreme, taking the longest to have a one-percentage point increase, here being 2.91 to 3.89 years (Table 2).

Smaller disciplines of COMP and PHYS—compared with huge disciplines of MED and BIO—look stagnant, with ENG changing its gender composition at a little higher rate but with a very low starting point. In 1990, the percentage of women scientists in ENG was by far the lowest of all disciplines (5.61%), still being the lowest in 2023, despite increasing almost threefold (to 15.88%).

In aggregate terms of disciplines, Europe does not differ much from its four selected comparator countries, except for Japan, which has traditionally been a global leader in the lowest participation of women in science, together with South Korea (Kwiek and Szymula, 2023). (The subsample of scientists from the four countries are characterized in Supplementary Table 3, and the details of regression model statistics for each country separately are shown in Supplementary Table 5).

**Table 1 Percentage of publishing nonoccasional STEMM women scientists in European countries in selected years from 1990 to 2023 by discipline (in %) ($N = 1{,}740{,}985$) (For the four comparator countries, see Supplementary Table 7).**

| Discipline | Percentage of women scientists in selected years | | | | | | | |
|---|---|---|---|---|---|---|---|---|
| | **1990** | **1995** | **2000** | **2005** | **2010** | **2015** | **2020** | **2023** |
| AGRI | 20.83 | 25.19 | 29.15 | 33.97 | 38.37 | 41.63 | 43.15 | 43.15 |
| BIO | 32.18 | 36.77 | 39.89 | 43.30 | 45.77 | 47.30 | 47.86 | 47.58 |
| CHEM | 18.82 | 22.61 | 26.43 | 30.89 | 33.65 | 34.88 | 35.25 | 35.07 |
| COMP | 11.62 | 12.19 | 13.77 | 14.27 | 16.60 | 17.13 | 18.86 | 19.00 |
| EARTH | 12.69 | 15.28 | 18.48 | 21.71 | 25.61 | 27.78 | 28.95 | 28.97 |
| ENG | 5.61 | 6.72 | 8.42 | 10.31 | 11.89 | 14.07 | 15.67 | 15.88 |
| ENVIR | 16.18 | 21.82 | 26.17 | 31.86 | 36.48 | 40.32 | 42.19 | 41.92 |
| IMMU | 34.10 | 38.45 | 42.18 | 46.94 | 51.80 | 54.26 | 55.69 | 55.60 |
| MATER | 16.89 | 19.80 | 24.53 | 28.20 | 32.01 | 32.50 | 33.28 | 33.42 |
| MATH | 10.93 | 12.35 | 16.03 | 17.94 | 19.58 | 20.53 | 21.25 | 21.27 |
| MED | 24.51 | 28.41 | 32.90 | 37.12 | 41.57 | 45.48 | 47.68 | 48.07 |
| NEURO | 31.06 | 33.42 | 36.54 | 40.25 | 42.19 | 44.51 | 45.64 | 44.59 |
| PHARM | 30.98 | 34.87 | 39.19 | 42.99 | 46.19 | 49.26 | 52.78 | 52.07 |
| PHYS | 9.64 | 11.27 | 13.34 | 15.13 | 16.45 | 17.40 | 18.05 | 18.29 |
| **TOTAL** | **22.44** | **25.77** | **29.00** | **32.66** | **35.78** | **38.40** | **39.98** | **40.23** |

**Table 2 Regression model statistics (selection): Trends in the percentage of publishing nonoccasional STEMM women scientists in European countries by discipline, 1990–2023 ($N = 1{,}740{,}085$) (Full statistics in Supplementary Table 4; for the four comparator countries, see Supplementary Table 5).**

| | Slope[a] | | | Intercept[a] | Quality measures | | Change |
|---|---|---|---|---|---|---|---|
| **Discipline** | **Value** | **LB** | **UB** | **Value** | **R2** | **Standard error** | **Time needed for a 1 p.p. change (in years)** |
| AGRI | 0.74 | 0.698 | 0.780 | 22.10 | 0.975 | 1.166 | 1.35 |
| BIO | 0.48 | 0.436 | 0.528 | 34.67 | 0.930 | 1.300 | 2.08 |
| CHEM | 0.54 | 0.481 | 0.600 | 20.81 | 0.908 | 1.685 | 1.85 |
| COMP | 0.26 | 0.243 | 0.271 | 10.94 | 0.977 | 0.387 | 3.89 |
| EARTH | 0.55 | 0.524 | 0.585 | 13.15 | 0.976 | 0.862 | 1.80 |
| ENG | 0.34 | 0.334 | 0.352 | 5.19 | 0.994 | 0.259 | 2.91 |
| ENVIR | 0.85 | 0.800 | 0.902 | 18.02 | 0.971 | 1.445 | 1.18 |
| IMMU | 0.71 | 0.662 | 0.751 | 35.61 | 0.968 | 1.255 | 1.42 |
| MATER | 0.55 | 0.492 | 0.607 | 18.46 | 0.916 | 1.632 | 1.82 |
| MATH | 0.34 | 0.307 | 0.381 | 11.84 | 0.912 | 1.046 | 2.91 |
| MED | 0.78 | 0.750 | 0.809 | 25.08 | 0.988 | 0.841 | 1.28 |
| NEURO | 0.48 | 0.440 | 0.514 | 31.73 | 0.955 | 1.015 | 2.10 |
| PHARM | 0.67 | 0.639 | 0.703 | 32.59 | 0.981 | 0.913 | 1.49 |
| PHYS | 0.28 | 0.256 | 0.294 | 10.32 | 0.962 | 0.538 | 3.63 |
| **TOTAL** | **0.58** | **0.555** | **0.610** | **23.24** | **0.983** | **0.746** | **1.72** |

[a]For slope and intercept, p-value in all cases <0.0001.

The participation of women in science in Europe is increasing in a manner comparable to the USA (with slope values for all disciplines combined of 0.58 and 0.60, respectively, but slower than in Canada and much slower than in Australia). The growth of women's participation in science in Europe and in the comparator countries is driven by the same three disciplines: AGRI, ENVIR, and MED. As clearly seen in Fig. 3, in the USA, Canada, and Australia, the clusters of countries in terms of growth patterns are exactly the same: BIO and MED, the two largest disciplines, as well as IMMU, showed high starting points back in 1990 and are still the leaders in women's participation; the four disciplines of COMP, ENG, MATH, and PHYS showed a very low starting point, very slow growth rate (as testified by slope values), and very low participation rate in 2023, and the highest time for a one-percentage point change in women participation over the period studied. Japan has emerged in the current study as an outlier, with very low starting levels and extremely low growth rates in the four disciplines and in CHEM.

**A cross-sectional view: the participation of women in 2023 by age groups**. In contrast to a trend account focused on change over time, a horizontal cross-sectional account indicates how men and women scientists were distributed within the disciplines examined at a single point in time. A closer look at the most recent year for which full data are available (2023) for all age groups combined shows huge cross-disciplinary differentiation, with the share of women in IMMU and PHARM being three times higher than the share of women in the highly mathematized fields of ENG, PHYS, COMP, and MATH (Fig. 4, Left).

In research and policy discussions, referring to "women in STEMM fields" in general seems useless because the differences within STEMM are too large to allow for general conclusions. Much more fine-grained studies based on the STEMM disciplines are needed to see the ongoing changes in their full (quantitative) complexity. There are several disciplines with the share of women close to or exceeding 50%, and there are four disciplines where the share of women is in the range of 15–21% (MATH 21.27%, COMP 19.00%, PHYS 18.29%, and ENG 15.88%), for which the emergent data-driven picture is fundamentally different.

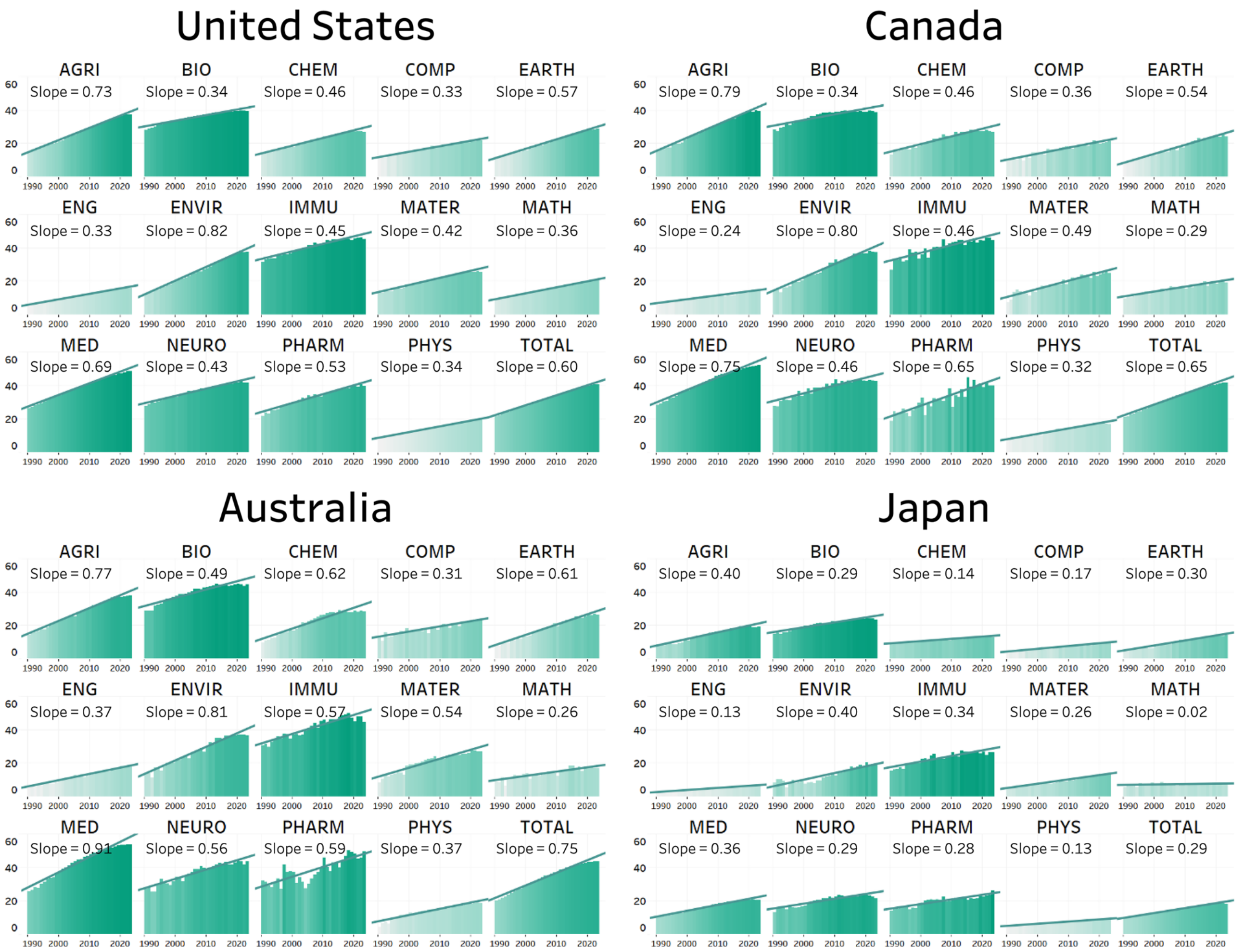

**Fig. 3** Growth in participation in publishing nonoccasional STEMM women scientists in the four comparator countries over time by discipline, 1990–2023 (in %).

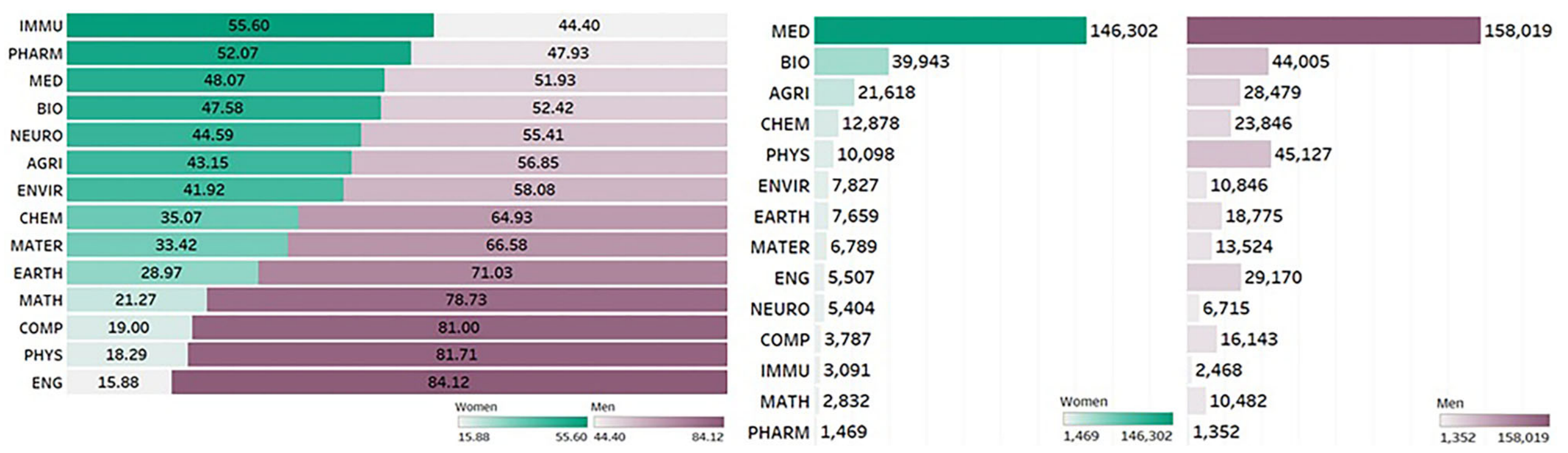

**Fig. 4** The percentage (Left) and numbers (Right) of publishing nonoccasional STEMM scientists in European countries by discipline and gender (in %) (row percentages: 100% horizontally), 2023 ($N = 684{,}155$, of which 275,204, or 40.23% identified as women).

To provide an example of using a high granularity level at our disposal, we know from our computations that, in ENG Engineering across Europe (in 2023), there were 5507 women publishing alongside 29,170 men (Fig. 4, Right). However, what we have not been aware of—before the advent of structured Big Data and new analytical possibilities they offered—is the exact distribution of men and women by age groups in ENG as well as in all other disciplines examined.

In simple terms, we are aware of men mathematicians (computer scientists, physicists, and astronomers, and engineers) being the vast majority among older mathematicians in our universities—and we are aware of increasing numbers of women scientists in these disciplines. However, it was not possible to show the patterns of disciplinary, gender, and generational changes at a large scale, beyond national borders without Big Data. What worked as an analytical strategy to analyze the inflow of women to science for individual countries (specifically, the USA) did not work globally or regionally without the new datasets.

**Table 3 The percentage of female scientists among publishing nonoccasional STEMM scientists in European systems by discipline in selected age groups (younger vs. older), 2023.**

| | Percentage of women scientists in selected age groups (younger and older) | | | | | |
|---|---|---|---|---|---|---|
| | 5 or less years | 6–10 years | 31–35 years | 36–40 years | 41–45 years | 46 & more years |
| IMMU | 67.34 | 63.19 | 39.56 | 28.73 | 22.13 | 19.15 |
| PHARM | 58.74 | 56.22 | 41.11 | 35.35 | 34.00 | 30.00 |
| NEURO | 57.95 | 52.72 | 31.45 | 26.51 | 28.64 | 23.58 |
| MED | 56.57 | 54.14 | 33.19 | 26.67 | 22.37 | 18.14 |
| BIO | 56.31 | 52.04 | 37.53 | 34.00 | 29.20 | 22.05 |
| AGRI | 54.19 | 49.23 | 29.91 | 24.88 | 21.65 | 11.81 |
| ENVIR | 48.06 | 46.77 | 30.82 | 20.70 | 17.60 | 7.95 |
| CHEM | 39.09 | 35.94 | 28.27 | 23.85 | 19.72 | 14.80 |
| MATER | 37.02 | 34.65 | 22.48 | 22.43 | 21.09 | 7.96 |
| EARTH | 36.16 | 34.60 | 19.86 | 17.23 | 14.30 | 8.94 |
| PHYS | 22.92 | 21.27 | 13.98 | 12.17 | 12.04 | 7.86 |
| ENG | 21.44 | 19.12 | 9.91 | 7.64 | 6.51 | 5.11 |
| MATH | 21.27 | 21.02 | 17.49 | 13.93 | 12.55 | 7.51 |
| COMP | 19.98 | 19.67 | 16.24 | 13.75 | 11.44 | 6.85 |
| **TOTAL** | **49.15** | **45.30** | **28.52** | **23.90** | **20.15** | **14.80** |

A horizontal approach (i.e., percentage shown is the percentage of women in a given age cohort; all men and women scientists from an age cohort in a discipline = 100%) (For the four comparator countries, see Supplementary Table 8)

Our refined analysis focuses on generations. In disciplines, as well as in countries, institutions, and their departments, there is always a mix of male and female scientists from different age groups. Our research interest is how this generational mix is changing over time from a cross-disciplinary perspective in terms of numbers and proportions of young and old women.

We analyze the participation of women scientists among younger or early-career scientists (from younger age groups) and among older or late-career scientists (from older age groups, Table 3). Older scientists—with a longer publishing experience—have been in the science publishing system for decades, and younger scientists—with shorter publishing experience—have been participants in the system for just a few years. Structured Big Data provide us with the full details of their participation in science: They leave specific digital traces in all their indexed publications.

Importantly, the current increased share of women scientists by generation (2023) indicates their increasing inflow into disciplines over the past 30 years and more (i.e., women scientists in the old age groups today formed young age groups in the past when they started publishing). Academic age is determined on the basis of the date of the first indexed publication. All scientists publishing in academic science in 2023 for some time are actually survivors in science, with survival time ranging from several years to several decades (using Kaplan–Meier survival analysis, we have examined gender differences in attrition and retention rates for the 2000–2021 period for 38 OECD countries by STEMM discipline elsewhere; see Kwiek and Szymula, 2024).

The percentage of women scientists among publishing nonoccasional STEMM scientists in European systems by discipline in selected age groups in 2023 shows the power of structured Big Data: Although for scientists in the academic age group 41–45 years the percentage of women for all disciplines combined is 20%, for the youngest age group, it is almost 50% (20.15%, 49.15%, respectively, Table 3). At the very same point in time (2023), in six STEMM disciplines, the majority of the youngest publishing scientists (presumably still in doctoral schools) are women, and in five, women are the majority in the age group of 6–10 years (IMMU, PHARM, NEURO, and the two largest in STEMM: MED and BIO).

A graphical presentation of our generational analysis clearly indicates (see Fig. 5) that there are disciplines where the mix of men and women scientists rapidly changes by age group (as in AGRI, MED, and IMMU), and those where the mix is changing very slowly (COMP, MATH, and PHYS). Here, it is enough to look at the shapes of the current (academic) age structure and keep in mind that the current gender structure in disciplines reflects previous inflows of men and women into disciplines. Women already form more than 50% of scientists in six STEMM disciplines out of 14: AGRI, BIO, IMMU, MED, NEURO and PHARM. However, the data show entry-level parity only and the parity is not reflected in more senior-level positions, e.g., in generations with at least 26 years of publishing experience where access to leadership positions is reported to be limited and research funding success is reported to be lower for women.

If we treat the current snapshot of survivors in science (2023) as a crude insight into the changing gender dynamics in the past 30 years, the findings are clear: There are two extremes, and at one of them, the changes in disciplines are marginal (in such disciplines as COMP and MATH, comparing age groups 6–10 years and 26–30 years: 19.67% and 17.25% in COMP, and 21.02% and 22.32% in MATH); and at the other extreme, the changes in disciplines are substantial, driving the participation of women beyond 55–60% for the young group (IMMU and PHARM).

The changing mix of scientists by gender has practical implications. The isolation of women scientists in the STEMM disciplines, specifically in the four math-intensive disciplines, has been steadily decreasing. Using our micro-level data on individuals, we can contrast the staggering isolation of women in older generations (e.g., women currently publishing for 41–45 years), with very high visibility—reaching 50% and more—of women in younger generations (e.g., women currently publishing for 5–10 years) within the same disciplines in the same year (2023).

In some disciplines, young women scientists are the majority of young scientists, but the share of older women scientists is only about a quarter of older scientists (with MED and NEURO as prime examples). In the four math-intensive disciplines, at the other extreme, the share of women for the younger age groups is about 20%, and the shares for the older age groups are in the range of 11–13% (and below 10% for ENG).

As our data show (see Table 4), in Europe in ENG in 2023, there were 1461 (nonoccasional, publishing) women vs. 6180 men in the young age group of 6–10 years—as opposed to merely 79 women vs. 955 men in the older age group of 36–40 years (and 12 vs. 223 in the oldest age group examined, 46 years and more), which meant their shares were at the levels of 19.12%, 7.64%, and 5.11%, respectively. From a generational perspective, the transition in ENG means changing from work settings (mostly universities) with a 20-times difference in participation between men and women in the past decades to settings with a five-times difference today. Arguably, this is a sea of change.

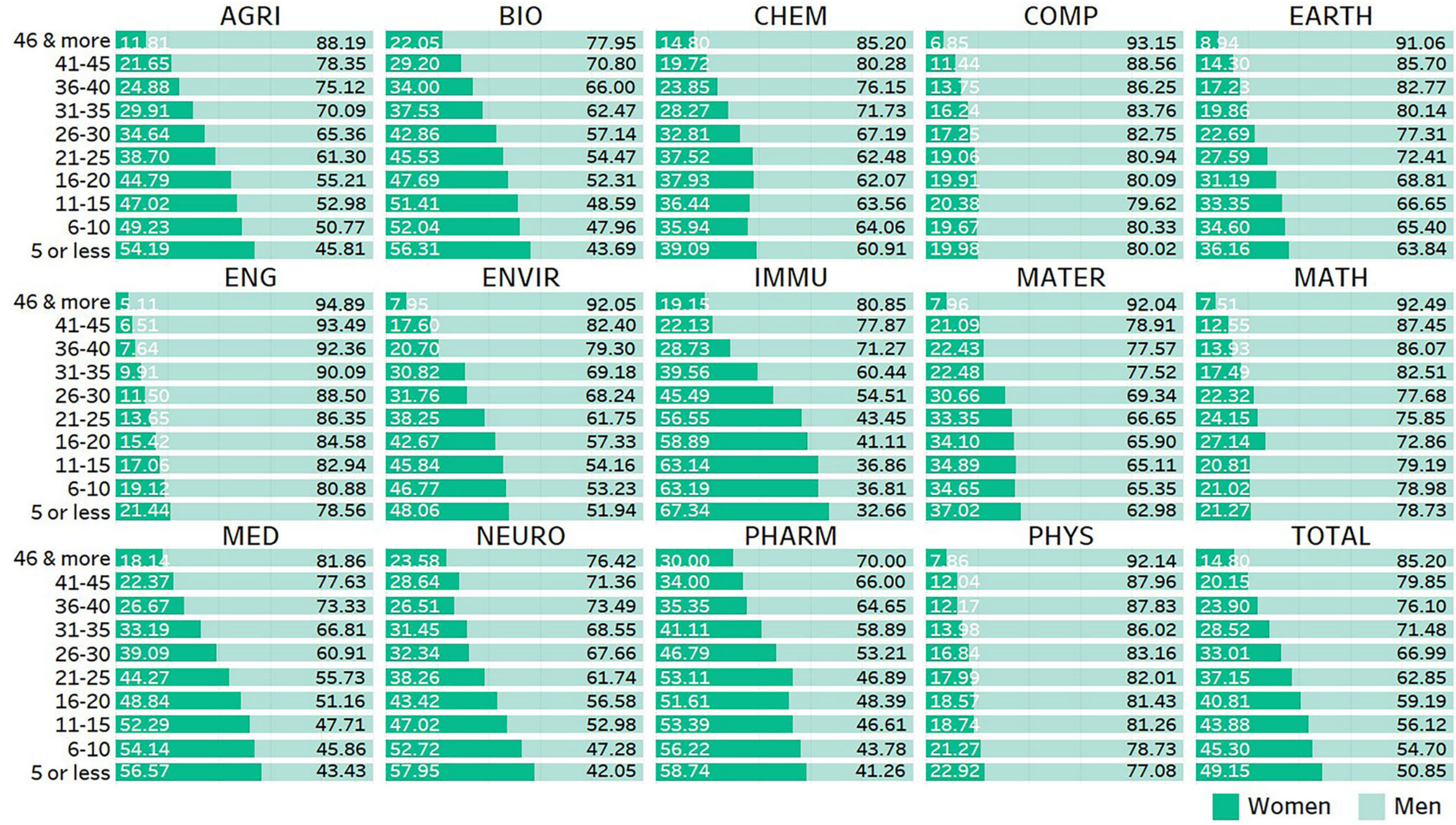

**Fig. 5 Participation of women in science by academic generations.** Horizontal approach: distribution of publishing nonoccasional STEMM scientists in European systems by discipline, age group, and gender (row percentages: 100% horizontally), 2023 ($N = 684{,}155$).

**Table 4 The number of men and women scientists among publishing nonoccasional STEMM scientists in European systems by discipline in selected age groups, 2023.**

| Discipline | Number of men and women scientists in selected age groups (young and old) | | | | | | |
|---|---|---|---|---|---|---|---|
| | **Gender** | **5 or less years** | **6–10 years** | **31–35 years** | **36–40 years** | **41–45 years** | **46 & more years** |
| AGRI | Women | 2792 | 4794 | 926 | 403 | 160 | 32 |
| | Men | 2360 | 4944 | 2170 | 1217 | 579 | 239 |
| BIO | Women | 4737 | 9223 | 2137 | 1183 | 499 | 174 |
| | Men | 3675 | 8500 | 3557 | 2296 | 1210 | 615 |
| CHEM | Women | 2034 | 3035 | 679 | 346 | 153 | 58 |
| | Men | 3169 | 5410 | 1723 | 1105 | 623 | 334 |
| COMP | Women | 250 | 702 | 208 | 74 | 27 | 5 |
| | Men | 1001 | 2867 | 1073 | 464 | 209 | 68 |
| EARTH | Women | 699 | 1695 | 382 | 205 | 109 | 27 |
| | Men | 1234 | 3204 | 1541 | 985 | 653 | 275 |
| ENG | Women | 756 | 1461 | 172 | 79 | 35 | 12 |
| | Men | 2770 | 6180 | 1563 | 955 | 503 | 223 |
| ENVIR | Women | 1079 | 1958 | 266 | 95 | 41 | 7 |
| | Men | 1166 | 2228 | 597 | 364 | 192 | 81 |
| IMMU | Women | 369 | 678 | 163 | 77 | 27 | 9 |
| | Men | 179 | 395 | 249 | 191 | 95 | 38 |
| MATER | Women | 1199 | 1741 | 194 | 109 | 58 | 9 |
| | Men | 2040 | 3284 | 669 | 377 | 217 | 104 |
| MATH | Women | 218 | 502 | 174 | 94 | 62 | 22 |
| | Men | 807 | 1886 | 821 | 581 | 432 | 271 |
| MED | Women | 24,465 | 36,102 | 5775 | 2875 | 1198 | 448 |
| | Men | 18,785 | 30,585 | 11,626 | 7904 | 4158 | 2021 |
| NEURO | Women | 656 | 1504 | 189 | 110 | 59 | 25 |
| | Men | 476 | 1349 | 412 | 305 | 147 | 81 |
| PHARM | Women | 252 | 366 | 74 | 35 | 17 | 9 |
| | Men | 177 | 285 | 106 | 64 | 33 | 21 |
| PHYS | Women | 1303 | 2356 | 632 | 320 | 191 | 75 |
| | Men | 4383 | 8719 | 3889 | 2310 | 1395 | 879 |
| **TOTAL** | **Women** | **40,809** | **66,117** | **11,971** | **6005** | **2636** | **912** |
| | **Men** | **42,222** | **79,836** | **29,996** | **19,118** | **10,446** | **5250** |

Currently research-active older women in ENG in Europe were working in their institutions (scattered across European countries) at times when 92–95% of their colleagues were men. This was a huge challenge as survey- and interview-based studies have been pointing out for decades. The changing mixes of men and women scientists by academic age group in Europe have been similar in the changes in the four comparator countries (Supplementary Fig. 1 and Supplementary Fig. 2).

## Discussion and conclusions

Although traditionally the growth of science has been estimated through the growth of publications and scientists (no gender specified), new structured Big Data of the bibliometric type combined with gender-detecting tools allow us to estimate the contribution of publishing scientists as men and women scientists (with specified gender) to the growth of science by disciplines and over time. Our detailed data at the micro level of individual scientists show a substantial inflow of women into science in the past three decades (1990–2023), which is powerfully differentiated by STEMM disciplines.

Our research based on the microdata derived from the raw Scopus dataset follows a recent line of large-scale bibliometric studies produced mostly in the emergent field of science of science and focused on women in science from a historical perspective of the past few decades. Large-scale studies which are parallel to our study are focused on various aspects of gender disparities in academic science which previously had been examined only in smaller scale research, mostly through interview research and survey research, often limited to individual institutions.

Large-scale studies discuss credit allocation in science or who tends to become an author and who tends to be denied this opportunity from among mixed-sex team members (Ross et al., 2022); publishing career lengths and drop-out rates or who tends to leave the academic profession earlier and who tends to leave it later on in academic careers (Huang et al., 2020); gendered nature of authorship or who tends to be in more the prestigious first and the last author positions in academic publishing and who tends to be in less prestigious middle author positions (Ni et al., 2021); gender imbalances among top-cited scientists or who tends to be heavily cited (Ioannidis et al., 2023); gender differences in self-citations or who tends to self-cite more and who tends to self-cite less (King et al., 2017); the role of gender in scholarly authorship or who is overrepresented in single-authored papers (West et al., 2013); and the unequal impact of parenthood in academia or who tends to pay higher productivity penalty for parenthood and who does not pay it at all (Morgan et al., 2021). The above large-scale research uses three types of methodological approaches and data sources: bibliometric data (as in King et al., 2017; Ioannidis et al., 2023; Huang et al., 2020; West et al., 2013; and Lariviere et al., 2013), researcher surveys (as in Ross et al., 2022; and Morgan et al., 2021), and follow-up interviews with survey participants (as in Ross et al., 2022; and Ni et al., 2021). This research follows broadly conceptual frameworks used in studying scientific research careers (Canibano et al., 2019).

None of the papers mentioned above referred in detail to increasing women's participation in science but all of them are relevant when gender disparities in science are discussed. And participation is closely linked to gender disparities: as participation increases, reaching the levels of gender parity (50/50) or gender balance (40/60), gender disparities tend to decrease. Our research theme of diversified and increasing access to the publishing academic profession for women in the past three decades is clearly related to the theme of gendered nature of authorship because authorship is the primary form of symbolic capital in science and women express concerns regarding the fair attribution of credit in both a survey reported and follow-up interviews (Ni et al., 2021). A survey of 103,296 researchers shows that there are gender differences in disagreements about authorship. A study of eight million papers across the natural sciences, social sciences and humanities shows that men predominate in the first and last author positions and women are significantly underrepresented as authors of single-authored papers (West et al., 2013).

Our theme of increasing participation of women in science is also linked to the theme of parenthood which slows down or stops entirely academic careers of women – and not careers of men. A survey of 3064 tenure-track faculty in the United States and Canada shows that parenthood explains most of the gender productivity gap by lowering the average short-term productivity of mothers (Morgan et al., 2021). Also related to our research theme, Huang and colleagues (2020) in their study of over 1.5 million authors demonstrate that there are two gender invariants found in their bibliometric research: men and women publish at a comparable annual rate and have equivalent career-wise impact for the same size body of work. Gender differences in productivity and impact can be explained by differences in publishing career lengths and drop-out rates (on gender gaps, see Kwiek and Roszka, 2022a).

Participation in science is also clearly linked to being credited in science. Ross et al. (2022) show that women in research teams are significantly less likely than men to be credited with authorship–women are significantly less likely to be named on a given article or patent produced by their team relative to their male peers. Being not credited for doing science certainly increases probability of leaving science – the chances of attrition for women in academic science increase. Finally, again with long-term implications for academic careers, men tend to have much higher self-citation rates than women, which is important for scholarly visibility and cumulative advantage in academic careers (King et al., 2017). In the past two decades studied, men self-cited 70 percent more than women and each self-citation leads to three additional external citations, with possible implications for promotion prospects (Fowler and Aksnes, 2007).

Women's participation in science is closely related to the various gender gaps in science and European science studied in this paper is not an exception. Productivity gap and citations gap may increase the chances for attrition from academic science, as may do collaboration gap, funding gap, and promotion gap. Women come to science in generally lower proportions than men and they leave science not only in higher proportions than men but also earlier in their careers than men (Kwiek and Szymula, 2024).

Different data-driven stories about growth in science as driven by women emerge when current data and historical trends (1990–2023) are explored from a gender and academic age group perspective. Our study shows a powerfully divided picture: There are disciplines with the share of women close to or exceeding 50%, which is a classical aim of gender equity in science; and there is a cluster of four disciplines where the share of women is still in the range of 15–21% (2023: MATH 21.27%, COMP 19.00%, PHYS 18.29%, and ENG 15.88%), for which the emergent data-driven picture is fundamentally different.

Our study shows that a traditional (somehow) monolithic segment of "STEMM disciplines" emerges from this research as divided between the disciplines in which women's share rose most rapidly and the disciplines in which the role of women was still marginal.

At one extreme, there are disciplines in which about 50% of currently publishing scientists are women (BIO, IMMU, PHARM, and MED, the largest discipline studied, Table 1); there are disciplines in which more than 50% of currently young scientists (with less than 10 years of publishing experience) are women (IMMU, PHARM, NEURO, and the two largest disciplines studied, MED and BIO, Table 3). These are disciplines where gender parity (50/50) and gender balance (40/60) has already been achieved among all research-active and young research-active scientists. No matter how welcoming (or unwelcoming) these disciplines have been in the past decades, a massive inflow of women as the drivers of growth is a fact of life across the 32 European systems examined. Similar patterns were found for the USA, Canada, and Australia (but not for Japan), using exactly the same methodological approach. In the USA, 45.21% of scientists publishing in 2023 in IMMU and 48.34% publishing in MED were women scientists, the average for all STEMM disciplines combined being 40.51%, which is nearly equal to the European average of 40.23%. More than 50% of young American scientists (with 5 or less years of publishing experience) in four STEMM disciplines (IMMU, NEURO, PHARM, and MED) were women. Similarly, more than 60% of young publishing scientists in MED in Canada and Australia were women.

At the other extreme, there is a cluster of highly mathematized disciplines (MATH, COMP, ENG, and PHYS) in which the growth of science is only marginally driven by women. The participation of women is low, and the rate of growth is small, with trends indicating no major changes in the past (and possibly in the near future). Graphically, in the past three decades, these are just different universes (Fig. 2)—both in Europe and in the four comparator countries. For reasons not discussed in the present paper but speculated and researched in a long array of publications, women tend not to enter the four disciplines as greatly as the other 10. (In a recent cohort-based study, we show that once women start publishing, their retention rate in these four disciplines is not lower than men's, as opposed to all other disciplines (Kwiek and Szymula, 2024).

The growth rates for women scientists show the inflow of women into science from a temporal perspective. There are fast growers and laggards, and trend data indicate the direction of changes in different clusters of STEMM disciplines. The growth in the numbers and shares of women seems far from saturation—except for the four highly mathematized disciplines where the growth is slow and women's participation, compared with all other disciplines, is very low (as shown by the differentiation of slopes in regression model statistics of trends in the percentage of publishing women scientists, with a threefold difference between COMP and AGRI, ENVIR, and MED, Table 2).

The policy issues that emerge from the present research should be related to keeping women who entered science in the past decades in knowledge production sectors (academia and beyond); and should be related to increasing women's participation in those disciplines where it is low and for which growth rates are marginal. Both the gender participation gap was studied (Huang et al., 2020; Larivière et al., 2013; Ni et al., 2021; West et al. 2013), and the gender attrition gap was explored (Spoon et al., 2023; Xu, 2008). However, the fundamental difference within the STEMM fields was not emphasized strongly enough. In other words, the problems of participation and retention/attrition in science from a gender perspective are acute in only some disciplines; in others, as our current and trend data show, they are barely overcome.

In research and policy discussions, referring to "women in STEMM fields" in general, without further specifications, and merely in a general contrast to women in non-STEMM fields (especially ECON and PSYCH in social sciences), does not seem to be fruitful anymore. Intradisciplinary differences within STEMM are too large to allow the generation of general conclusions or advising general policy interventions.

Fine-grained studies based on STEMM disciplines—as well hundreds of subdisciplines and thousands of topic clusters—are needed to see the dynamic of ongoing changes in the growth of science in its full (quantitative) complexity.

As speculation only, we can calculate when an increase to a minimum of 40% participation rate for women (indicating a gender balance) can be achieved for any discipline below this level—combining the level of participation in 2023 (Table 1) and the time needed for a one-percentage point increase in participation-based trend lines (Table 2)—should current trends continue undisturbed. Gender balance (40/60) has already been achieved in half of STEMM disciplines (including the two largest, MED, and BIO), and in all STEMM disciplines combined.

Finally, the application of bibliometric datasets and Big Data analytics to the estimation of the growth of science from a gender perspective comes at a cost, introducing new challenges and requiring methodological choices. Therefore, we have discussed in detail the major challenges in changing the unit of analysis from the publication to the scientist; from the scientist (no gender specified) to man or women scientist; and from single-nation research to multicountry or global research into the growth of science based on individuals. The critical points in this type of research are gender determination, academic age determination, and discipline determination at the micro level of individuals.

In the present research, we assumed that the future growth of science and of its workforce (by gender) is hidden behind the current participation trends in science (by gender). The massive inflow of women to STEMM fields has been observed in the past three decades, and a massively increasing presence of publishing women is observed among the youngest generations of scientists in the STEMM fields. Therefore, our current data on participation in science by academic generations and gender—from the youngest to the oldest men and women—indicate the general direction of future changes in the science enterprise, with European science being defined in the present paper as a testing ground for the use of our empirical strategy. However, in the current research, we are not involved in predictive analytics about the future because the number of data points in the past has been too limited to present accurate long-time trend analyses.

### Data availability

We used data from Scopus, a proprietary scientometric database. For legal reasons, data from Scopus received through collaboration with the ICSR Lab owned by Elsevier cannot be made openly available.

## Acknowledgements

M.K. gratefully acknowledges the support provided by the grant from MNISW (NDS grant no. NdS-II/SP/0010/2023/01). L.S. is grateful for the support of his doctoral studies provided by the NCN grant 2019/35/0/HS6/02591. We gratefully acknowledge the assistance of the International Center for the Studies of Research (ICSR) Lab with Kristy James and Alick Bird. We also want to thank Dr. Wojciech Roszka from the CPPS Poznan Team for many fruitful discussions.

## Author contributions

Marek Kwiek: Conceptualization, Data curation, Formal analysis, Investigation, Methodology, Resources, Software, Validation, Writing—original draft, Writing—review & editing. Lukasz Szymula: Conceptualization, Data curation, Formal analysis, Investigation, Methodology, Software, Validation, Visualization, Writing—original draft, Writing—review & editing.

## Competing interests

The authors declare no competing interests.

## Ethical approval

This article does not contain any studies with human participants performed by any of the authors.

## Informed consent

This article does not contain any studies with human participants performed by any of the authors.

## Additional information

**Correspondence** and requests for materials should be addressed to Marek Kwiek.

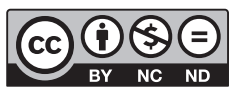